\documentclass{book}    
\usepackage{piers}  
\begin{document}

\title{ On the Validity of Physical Optics  for Narrow-band  Beam Scattering and Diffraction   from the   Open Cylindrical Surface}
\maketitle

\author      {F. M. Lastname}
\affiliation {University}
\address     {}
\city        {Boston}
\postalcode  {}
\country     {USA}
\phone       {345566}    
\fax         {233445}    
\email       {email@email.com}  
\misc        { }  
\nomakeauthor

\author      {F. M. Lastname}
\affiliation {University}
\address     {}
\city        {Boston}
\postalcode  {}
\country     {USA}
\phone       {345566}    
\fax         {233445}    
\email       {email@email.com}  
\misc        { }  
\nomakeauthor

\begin{authors}

{\bf Shaolin Liao}    \\
Electrical and Computer Engineering, 1415 Engineering Drive, Univ.
of Wisconsin, Madison, U.S.A., 53706
\end{authors}
\begin{paper}

\begin{piersabstract}
The exact formulas for the induced electric surface current  (in the
scattering phenomenon) and the equivalent  electric surface current
(in the diffraction phenomenon) on the open cylindrical   surface
due to an arbitrary  narrow-band beam
  have been shown in their
closed-form expressions within the context of the cylindrical
harmonics, which gives information about the validity of the
Physical Optics (PO) approximation. Both the Electric Field Integral
Equation (EFIE) and the Magnetic Field Integral Equation (MFIE) are
used to find the induced (equivalent) electric surface currents in
the context of the cylindrical harmonics. The numerical example of
the scattering and diffraction of the Hermite Gaussian beam   from
the open cylindrical surface is shown. The result is useful for the
evaluation of the validity of the PO approximation in the
cylinder-like surface.
\end{piersabstract}


  \begin{center}
  {\bf I. Introduction}
\end{center}

 The Physical Optics (PO) approximation has been extensively used as
 the  approximation of the exact solution in many applications \cite{liao_image_2006, liao_near-field_2006, shaolin_liao_new_2005, liao_fast_2006,  liao_beam-shaping_2007, liao_fast_2007, liao_validity_2007, liao_high-efficiency_2008, liao_four-frequency_2009, vernon_high-power_2015, liao_multi-frequency_2008, liao_fast_2007-1, liao_sub-thz_2007, liao_miter_2009, liao_fast_2009, liao_efficient_2011, liao_spectral-domain_2019}, which include  microwave
imaging, reflector antenna design, and evaluation of Radar Cross
Section (RCS) \cite{Lin_1993, Schlobohm_1992, Hestilow_2000, Shaolin_conf}. It is
helpful to have an analytical formula to predict the behavior of the
PO approximation in order to use it effectively. In this article,
the exact closed-form expressions will be shown for the induced
(equivalent) electric surface currents on the open cylindrical
surface, from which the information of the validity of the PO
approximation is obtained for the cylinder-like surface. The scheme
used to illustrate the problem is given in Fig. \ref{scheme}. The
time dependence $e^{i \omega t}$ ($i=\sqrt{-1}$) has been assumed in
this article.

  \begin{center}
  {\bf II. The Cylindrical Harmonics}
\end{center}

The cylindrical modal expansion of the vector potential  $  { \bf
A}({\bf r})$ for the   electric surface current ${ \bf J}_s ({\bf
r}')$ on an arbitrary surface  in the cylindrical coordinate is
given as

 \vspace*{-0.1in}

{ \small

\begin{eqnarray}\label{FMGreen_cyl}
    { \bf A}  ({\bf r})    = \mu  \int\!\!\int_{S}
 \left[ \hspace{-0.07in} \begin{array}{cccc}  \\  \\  \end{array} g({\bf r- r' })
   { \bf J} _s  {\bf (r')} \hspace{-0.07in} \begin{array}{cccc}  \\  \\  \end{array} \right]   dS' = \frac{ \mu}{
i 8 \pi}  \int \! \! \int_{S}   \left[   \hspace{-0.07in}
\begin{array}{cccc}  \\  \\  \end{array}  { \bf J}   _s {\bf (r')}
    \int_{-\infty}^{\infty}      H_0^{(2)} \left( \Lambda \left|\boldsymbol{\rho}-\boldsymbol{\rho}'\right| \right)
    e^{ - i
h (z-z')} d h \hspace{-0.07in} \begin{array}{cccc}  \\  \\
\end{array}  \right]   dS'
 \end{eqnarray}
}

\hspace*{-0.22in}where  $\mu$ is the permeability of the homogeneous
medium. $H_0^{(2)}( \ {\cdot} \ )$ is Hankel function of the second
kind of order  0.   The scalar Green's function $g( \ \cdot \ )$ and
the transverse wave vector $\Lambda$ are defined as

 \vspace*{-0.1 in}

{  \small
\begin{eqnarray}\label{Green_cyl}
   g( \ {\cdot} \
) = \frac{e^{-i k | \ {\cdot} \ |}}{4 \pi | \ {\cdot}  \ |}, \ \
\Lambda =\sqrt{k^2-h^2}.
\end{eqnarray}
}

According to the cylindrical addition theorem,

\vspace*{-0.1in}

{ \small

 \begin{eqnarray}\label{Addition_cyl}
 \hspace*{-0.1in} H_0^{(2) } \left( \Lambda \left| \boldsymbol{\rho}-\boldsymbol{\rho}' \right| \right)
= \sum_{m=-\infty}^\infty \left\{\begin{array}{ccc} \left.
 H_m^{(2)} (
\Lambda \rho ) J_m(\Lambda \rho')    e^{i m ( \phi'  - \phi )}\right|_{\rho>\rho'}  \\
\\
\left.  J_m ( \Lambda \rho )  H_m^{(2)}(\Lambda \rho')    e^{i m (
\phi'  - \phi )}\right|_{\rho<\rho'}
\end{array} \right.  \hspace{0.2in}
\end{eqnarray}
}

\hspace*{-0.22in}where $\rho \equiv |\boldsymbol{\rho}| $ is the
observation coordinate and $\rho' \equiv |\boldsymbol{\rho}'| $ is
the source coordinate. $J_m ( \ {\cdot} \ )$ is Bessel function of
the first kind of integer order $m$   and $H_m^{(2)} ( \ {\cdot} \
)$ is Hankel function of  the second kind  of integer order $m$.
Substituting (\ref{Addition_cyl}) into (\ref{FMGreen_cyl}), the
cylindrical modal expansion of ${  { \bf A} ( {\bf r})}$   is
obtained,

 \vspace*{-0.2in}

{  \small
\begin{eqnarray}\label{Ag}
  { \bf A}^>_< ({\bf r})   & = &   \hbox{IFT}
   \left(     { \bf g}^{ >}_{   < } (m, h) \begin{array}{cc}  H_m^{(2)} ( \Lambda \rho ) \\
  J_m  ( \Lambda \rho ) \end{array}   \hspace{-0.12in} \begin{array}{cccc}  \\  \\  \end{array}  \right)
  \end{eqnarray}

   \vspace*{-0.2in}

\begin{eqnarray}
 \ \ \ \ \ \ \ \  { \bf g}^{   >}_{  < } (m, h)  =   \frac{ \mu}{i4}
  \int\!\!\int_{S}
  \left[    \begin{array}{cc}  J_m(\Lambda
\rho')    \\   H_m^{(2)} ( \Lambda \rho' )  \end{array}  { \bf J}_s
{\bf (r')}  e^{i ( m \phi' + h z' )}       \hspace{-0.07in}
\begin{array}{cccc}  \\  \\  \end{array}      \right] dS'  \nonumber
\end{eqnarray}

}

\hspace*{-0.22in}where, the the superscript ``$>$" denotes $\rho >
\rho'$ and the subscript ``$<$" denotes $\rho < \rho'$.  The Inverse
Fourier Transform (IFT) is defined as

  \vspace*{-0.1in}

{  \small

\begin{equation}\label{IFT}
\hbox{IFT} \left(  \hspace{-0.0in} \begin{array}{ccc}  \\
\end{array}   {\cdot}  \hspace{-0.0in} \begin{array}{ccc}  \\
\end{array}    \right)  =   \frac{1}{2 \pi}
\sum_{m=-\infty}^\infty  \left\{  \int_{-\infty}^{\infty}   \left[    \left(  \hspace{-0.0in} \begin{array}{ccc}  \\
\end{array}   {\cdot}  \hspace{-0.0in} \begin{array}{ccc}  \\
\end{array}  \right)   e^{- i ( m \phi + h z )} \hspace{-0.07in} \begin{array}{ccc}  \\ \end{array}   \right]
 d h \hspace{-0.09in} \begin{array}{ccc}  \\ \\ \\ \end{array} \right\}.
\end{equation}
}

The electromagnetic field (${\bf E}, {\bf H} $) is given as

 \vspace*{-0.05in}

{ \small
\begin{equation}\label{EH}
 {\bf E}^>_< ( {\bf r}) = - i \omega { \bf A}^>_< ({\bf r}) + \frac{1}{i \omega \mu
 \epsilon} \nabla \left[\nabla \cdot { \bf A}^>_< ({\bf r}) \hspace{-0.11in} \begin{array}{ccc}  \\ \\   \end{array}
 \right], \ \ \ \  {\bf H}^>_< ( {\bf r}) = \frac{1}{\mu }  \nabla \times   { \bf A}^>_< ({\bf r})
\end{equation}

}

\begin{figure}\centering
 \includegraphics[width=2.9 in, height= 2.3in]{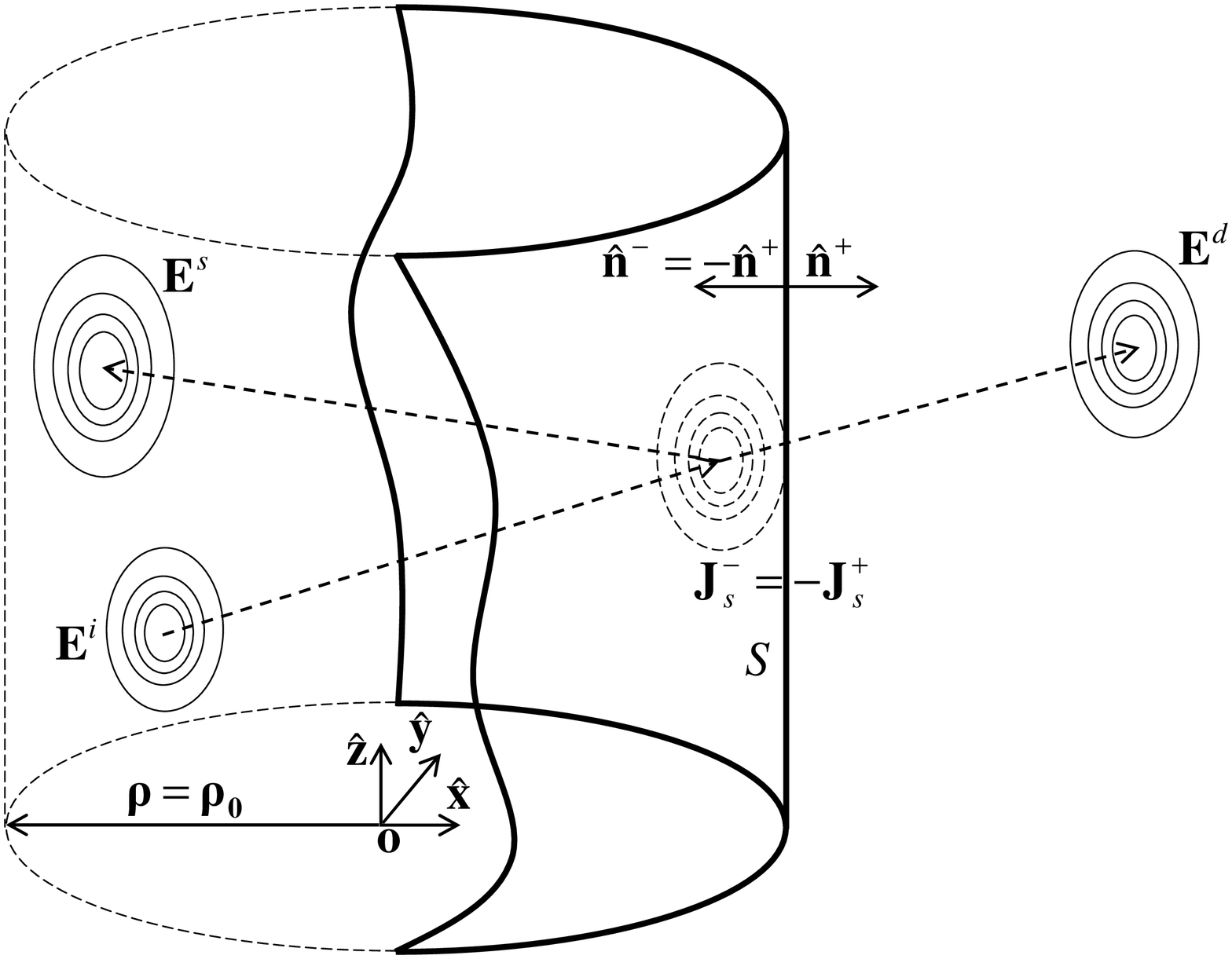}
\caption{The narrow-band beam  scattering and diffraction in the
cylindrical geometry: the incident field ${\bf E}^i$ propagates onto
cylindrical surface $S$ with  radius of $\rho_0$, then it could be
back-scattered to ${\bf E}^{s}$ if surface $S$ serves as a PEC
scatter, with induced surface current ${\bf J}_s^-$; or it may
forward-propagate  to ${\bf E}^{d}$ if it is a diffraction
phenomenon, with equivalent surface current ${\bf J}_s^+ = - {\bf
J}_s^-$. $\hat{\bf n}^+$ and $\hat{\bf n}^-$ are the outward and
 inward unit surface normals to $S$ respectively.
 }
 \label{scheme}
\end{figure}


  \begin{center}
  {\bf III. Exact Formulas for Induced and Equivalent Electric Surface Currents}
\end{center}

Due to the fact that ${\bf J}_s^- = -{\bf J}_s^+$ (see Fig.
\ref{scheme}), let's consider the scattering phenomenon and express
the incident electromagnetic  field  (${\bf E}^i, {\bf H}^i$) into
the cylindrical harmonics,

\vspace{-0.15in}

{ \small

\begin{eqnarray}\label{Ei}
 { \bf E}^i ( \rho )      =      \sum_{m=-\infty}^\infty
 \left\{
\int_{-\infty}^{\infty}    \left[  \hspace{-0.07in}
\begin{array}{ccc}  \\  \\  \end{array}  a^{h}_{m} \ {\bf M}^{
h }_{m} (\rho )   +  b^{ h}_{m}  \ {\bf N}^{h }_{m}
 (\rho ) \hspace{-0.07in} \begin{array}{ccc}  \\  \\  \end{array} \right] dh
  \hspace{-0.07in} \begin{array}{ccc}  \\  \\ \\  \end{array}
  \right\}
  \end{eqnarray}

\vspace{-0.1in}

\begin{eqnarray}\label{Hi}
 { \bf H}^i ( \rho )      =
\frac{i}{\eta}  \sum_{m=-\infty}^\infty
 \left\{
\int_{-\infty}^{\infty}    \left[  \hspace{-0.07in}
\begin{array}{ccc}  \\  \\  \end{array}  a^{h}_{m} \ {\bf N}^{
h }_{m} (\rho )   +  b^{ h}_{m}  \ {\bf M}^{h }_{m}
 (\rho ) \hspace{-0.07in} \begin{array}{ccc}  \\  \\  \end{array} \right] dh
  \hspace{-0.07in} \begin{array}{ccc}  \\  \\ \\  \end{array}
  \right\}
  \end{eqnarray}

  \vspace*{-0.2in}

\begin{eqnarray}\label{M}
   {\bf M}^{ h }_{m } ( {\bf r})   =
  \left[  \hspace{-0.07in}
\begin{array}{ccc}  \\  \\  \end{array} \hat{\boldsymbol{\rho}} \frac{m}{i\rho} H_m^{(2)}( \Lambda \rho)
   -   \hat{\boldsymbol{\phi}} \Lambda  \frac{\partial H_m^{(2)}( \Lambda \rho)}{\partial (\Lambda
   \rho)} \hspace{-0.07in}
\begin{array}{ccc}  \\  \\  \end{array} \right] e^{-i m\phi} e^{- i
h z}
\end{eqnarray}

 \vspace*{-0.1in}

\begin{eqnarray}\label{N}
   {\bf N}^{ h }_{m } ( {\bf r})   =
 \left[  \hspace{-0.07in}
\begin{array}{ccc}  \\  \\  \end{array}  \hat{\boldsymbol{\rho}} \frac{h \Lambda}{ik} \frac{\partial H_m^{(2)}( \Lambda \rho)}{\partial (\Lambda \rho)}
   -   \hat{\boldsymbol{\phi}} \frac{mh}{k\rho} H_m^{(2)}( \Lambda
   \rho) + \hat{\bf z} \frac{\Lambda^2}{k} H_m^{(2)}( \Lambda \rho) \hspace{-0.07in}
\begin{array}{ccc}  \\  \\  \end{array} \right] e^{-i m\phi} e^{- i
h z}
\end{eqnarray}

}

Similarly, express the scattered electromagnetic field (${\bf E}^s,
{\bf H}^s$) as

\vspace{-0.2in}

{ \small

\begin{eqnarray}\label{Es}
 { \bf E}^s ( \rho )      =      \sum_{m=-\infty}^\infty
 \left\{
\int_{-\infty}^{\infty}    \left[  \hspace{-0.07in}
\begin{array}{ccc}  \\  \\  \end{array}  c^{h}_{m} \ {\bf M}^{
h }_{m} (\rho )   +  d^{ h}_{m}  \ {\bf N}^{h }_{m}
 (\rho ) \hspace{-0.07in} \begin{array}{ccc}  \\  \\  \end{array} \right] dh
  \hspace{-0.07in} \begin{array}{ccc}  \\  \\ \\  \end{array}
  \right\}
  \end{eqnarray}

\vspace{-0.2in}

\begin{eqnarray}\label{Hs}
 { \bf H}^s ( \rho )      =
\frac{i}{\eta}  \sum_{m=-\infty}^\infty
 \left\{
\int_{-\infty}^{\infty}    \left[  \hspace{-0.07in}
\begin{array}{ccc}  \\  \\  \end{array}  c^{h}_{m} \ {\bf N}^{
h }_{m} (\rho )   +  d^{ h}_{m}  \ {\bf M}^{h }_{m}
 (\rho ) \hspace{-0.07in} \begin{array}{ccc}  \\  \\  \end{array} \right] dh
  \hspace{-0.07in} \begin{array}{ccc}  \\  \\ \\  \end{array}
  \right\}
  \end{eqnarray}

}

Now the induced electric surface current $  {\bf J}_s^-$ on the
cylindrical surface $S$ is given as

\vspace*{-0.1in}

{ \small

\begin{eqnarray}\label{Js}
 { \bf J}_s^- ( \rho_0)  =   \hat{\bf n}^- \times    \left[ { \bf H}^i ( \rho_0 ) + { \bf H}^s ( \rho_0
 )
 \right]         =       \left[ { \bf H}^i ( \rho_0 ) + { \bf H}^s ( \rho_0
 )
 \right]   \times \hat{\boldsymbol{\rho}}_0
   \end{eqnarray}

\vspace*{-0.2in}

\begin{eqnarray}
   =  \frac{ i  }{  \eta}  \sum_{m=-\infty}^\infty
 \left\{
\int_{-\infty}^{\infty}    \left[  \hspace{-0.07in}
\begin{array}{ccc}  \\  \\  \end{array} \left( a^{h}_{m} + c^{h}_{m} \right)    {\bf N}^{
h }_{m} (\rho_0 ) \times \hat{\boldsymbol{\rho}}_0   +   \left(
b^{h}_{m} + d^{h}_{m} \right)  {\bf M}^{h }_{m}
 (\rho_0 ) \times \hat{\boldsymbol{\rho}}_0   \hspace{-0.07in} \begin{array}{ccc}  \\  \\  \end{array} \right] dh
  \hspace{-0.07in} \begin{array}{ccc}  \\  \\ \\  \end{array}
  \right\}  \nonumber
  \end{eqnarray}

}


  \begin{center}
  {\bf 1. \small \ Electric Field Integral Equation (EFIE)}
\end{center}

Let's consider the TM mode (${\bf N}^{ h }_{m }$ for ${\bf E}$ and
${\bf M}^{ h }_{m }$ for ${\bf H}$) here. From (\ref{M})  and
(\ref{Js}),

\vspace*{-0.2in}

{ \small

\begin{eqnarray}\label{JsTM}
 { \bf J}_s^{-,\hbox{\tiny TM}} ( \rho_0)      =  \hat{\bf z}  \frac{ i  }{  \eta}   \sum_{m=-\infty}^\infty
 \left\{  \hspace{-0.07in} \begin{array}{ccc}  \\  \\ \\  \end{array}
\int_{-\infty}^{\infty}   \left( b^{ h}_{m} +  d^{ h}_{m} \right)
\Lambda  \left.\frac{\partial H_m^{(2)}( \Lambda \rho)}{\partial
(\Lambda
   \rho)}\right|_{\rho_0}  dh
  \hspace{-0.07in} \begin{array}{ccc}  \\  \\ \\  \end{array}
  \right\}
  \end{eqnarray}

  }

Substituting (\ref{JsTM}) into (\ref{EH}), the $z$-component of the
scattered electric field $  \hbox{   E}^{s,\hbox{\tiny TM}}_{z,<}  $
on the cylindrical surface is obtained ($\hbox{A}_{z,<}^{\hbox{\tiny
TM}} \equiv \hat{\bf z} \cdot { \bf A}_<^{\hbox{\tiny TM}}$),

\vspace*{-0.2in}

{ \small

\begin{eqnarray}\label{Eswhole}
 \hbox{   E}^{s,\hbox{\tiny TM}}_{<,z}(\rho_0) =   -i \omega
\left(\frac{\Lambda}{k} \right)^2  \hbox{A}_{z,<}^{\hbox{\tiny TM}}
( \rho_0)
\end{eqnarray}

\vspace*{-0.2in}

\begin{eqnarray}
  =    \frac{\pi \rho_0 }{i  2 k}
\sum_{m=-\infty}^\infty
 \left\{  \hspace{-0.07in} \begin{array}{ccc}  \\  \\ \\  \end{array}
\int_{-\infty}^{\infty}   \left( b^{ h}_{m} +  d^{ h}_{m} \right)
\Lambda^3  J_m( \Lambda \rho_0)   H_m^{(2)}( \Lambda \rho_0)
\left.\frac{\partial H_m^{(2)}( \Lambda \rho)}{\partial (\Lambda
   \rho)}\right|_{\rho_0}  dh
  \hspace{-0.07in} \begin{array}{ccc}  \\  \\ \\  \end{array}
  \right\} \nonumber
  \end{eqnarray}

}

Note that $  \hbox{   E}^{s,\hbox{\tiny TM}}_{z,<}  $ is given on
the whole cylindrical surface  that is just inside (infinitesimally
 close to)  cylindrical surface $S$, on both the
front side and the back side. It can be separated into two parts for
the  narrow-band beam, which can be seen from the property of Bessel
function,

\vspace*{-0.1in}

{ \small

\begin{eqnarray}\label{bessel}
J_m( \ \cdot \ )  =  \frac{ H_m^{(1)}( \ \cdot \ ) + H_m^{(2)}( \
\cdot \ )  }{2}
\end{eqnarray}
}

\vspace*{-0.1in}

The scattered electric field on the front side is thus given as

\vspace*{-0.2in}

 {\small
\begin{eqnarray}\label{Esfront}
\hbox{   E}^{s,\hbox{\tiny TM}}_{z,<,f}(\rho_0)  =    \frac{\pi
\rho_0 }{i 4 k} \sum_{m=-\infty}^\infty
 \left\{  \hspace{-0.07in} \begin{array}{ccc}  \\  \\ \\  \end{array}
\int_{-\infty}^{\infty}   \left( b^{ h}_{m} +  d^{ h}_{m} \right)
\Lambda^3  H_m^{(1)}( \Lambda \rho_0)   H_m^{(2)}( \Lambda \rho_0)
\left.\frac{\partial H_m^{(2)}( \Lambda \rho)}{\partial (\Lambda
   \rho)}\right|_{\rho_0}  dh
  \hspace{-0.07in} \begin{array}{ccc}  \\  \\ \\  \end{array}
  \right\}
  \end{eqnarray}

}

Now apply the EFIE on the cylindrical surface
$\hbox{E}^{s,\hbox{\tiny TM}}_{z,<,f} (\rho_0) = -
\hbox{E}^{i,\hbox{\tiny TM}}_z (\rho_0)$, from (\ref{Ei})   and
(\ref{Esfront}),

\vspace*{-0.1in}

{ \small

\begin{eqnarray}\label{bd}
b^{ h}_{m} +  d^{ h}_{m} =   \frac{2}{ \xi}b^{ h}_{m}, \ \ \ \ \ d^{
h}_{m} = \left[\frac{2 }{ \xi}-1\right] b^{ h}_{m}, \ \ \ \ \ \xi
\equiv i \frac{\pi}{2} \Lambda \rho_0 H_m^{(1)}( \Lambda \rho_0)
\left.\frac{\partial H_m^{(2)}( \Lambda \rho)}{\partial (\Lambda
   \rho)}\right|_{\rho_0}
  \end{eqnarray}

}

Note that $\xi \rightarrow 1$ for $\rho_0 \rightarrow \infty$, which
means that $ d^{ h}_{m} \rightarrow b^h_m$ and  the PO approximation
reduces to the exact induced electric surface current.


  \begin{center}
  {\bf 2. \small \ Magnetic Field Integral Equation (MFIE)}
\end{center}

Let's also take the TM mode (${\bf N}^{ h }_{m }$ for ${\bf E}$ and
${\bf M}^{ h }_{m }$ for ${\bf H}$) as an example. From (\ref{EH})
and  (\ref{JsTM}), the $\phi$-component of the scattered magnetic
field $ \hbox{ H}^{s,\hbox{\tiny TM}}_{\phi,<,f}(\rho_0)$  on the
front side of the cylindrical surface is found as

\vspace*{-0.2in} {\small
\begin{eqnarray}\label{Hsfront}
\hbox{  H}^{s,\hbox{\tiny TM}}_{\phi,<,f}  (\rho_0) =\frac{i}{\eta}
\sum_{m=-\infty}^\infty
 \left\{
\int_{-\infty}^{\infty}    \left[  \hspace{-0.07in}
\begin{array}{ccc}  \\  \\  \end{array}   \frac{\xi^\ast}{2} \left( b^{ h}_{m} +  d^{
h}_{m} \right)   \ {\bf M}^{h }_{m}
 (\rho ) \hspace{-0.07in} \begin{array}{ccc}  \\  \\  \end{array} \right] dh
  \hspace{-0.07in} \begin{array}{ccc}  \\  \\ \\  \end{array}
  \right\}
\end{eqnarray}
}

Now apply the MFIE on the cylindrical surface $ \hbox{
H}^{s,\hbox{\tiny TM}}_{\phi,<,f} (\rho_0)  +  \hbox{
H}^{i,\hbox{\tiny TM}}_\phi(\rho_0) = -
\hbox{J}_{s,z}^{-,\hbox{\tiny TM}}(\rho_0)$,

\vspace{-0.15in}

{ \small

\begin{eqnarray}\label{bdM}
b^{ h}_{m} +  d^{ h}_{m} =   \frac{2}{ 2- \xi^\ast}b^{ h}_{m}, \ \ \
\ \ d^{ h}_{m} =  \frac{ \xi^\ast }{2- \xi^\ast}   b^{ h}_{m}
  \end{eqnarray}
}

\vspace{-0.1in}

  It is not difficult to show that (\ref{bd})  and (\ref{bdM})
    are equivalent by using the Wronskian relation,

  \vspace{-0.15in}

{ \small

\begin{eqnarray}\label{Wronskian}
H_m^{(2)}( \Lambda \rho) \frac{\partial H_m^{(1)}( \Lambda
\rho)}{\partial (\Lambda
   \rho)} - H_m^{(1)}( \Lambda \rho) \frac{\partial H_m^{(2)}( \Lambda
\rho)}{\partial (\Lambda
   \rho)}  = \frac{i 4}{\pi \Lambda
   \rho }
  \end{eqnarray}

}


  \begin{center}
  {\bf \small 3. The Induced and Equivalent Electric Surface Currents}
\end{center}

Following the similar procedure, the induced electric surface
current for the TE mode (${\bf M}^{ h }_{m }$ for ${\bf E}$ and
${\bf N}^{ h }_{m }$ for ${\bf H}$) is given as

\vspace*{-0.2in}

{\small
\begin{eqnarray}\label{ac}
a^{ h}_{m} +  c^{ h}_{m} =   \frac{2}{ \xi^\ast }a^{ h}_{m}, \ \ \ \
\ c^{ h}_{m} = \left[\frac{2 }{ \xi^\ast}-1\right] a^{ h}_{m}
  \end{eqnarray}
}

Substituting (\ref{bd}) and (\ref{ac}) into (\ref{Js}), the total
induced and equivalent electric surface currents are obtained,

\vspace*{-0.2in}

{ \small

\begin{eqnarray}\label{Js_induced}
  { \bf J}_s^- ( \rho_0) = - { \bf J}_s^+ ( \rho_0)       =  \frac{ i 2 }{  \eta}  \sum_{m=-\infty}^\infty
 \left\{
\int_{-\infty}^{\infty}    \left[  \hspace{-0.07in}
\begin{array}{ccc}  \\  \\  \end{array} a_m^h  \frac{ {\bf N}^{
h }_{m} (\rho_0 ) \times \hat{\boldsymbol{\rho}}_0 }{\xi^\ast}     +
b_m^h \frac{{\bf M}^{h }_{m}
 (\rho_0 ) \times \hat{\boldsymbol{\rho}}_0 }{\xi}      \hspace{-0.07in} \begin{array}{ccc}  \\  \\
\end{array} \right] dh
  \hspace{-0.1in} \begin{array}{ccc}  \\  \\ \\  \end{array}
  \right\}
  \end{eqnarray}
}

From (\ref{Js_induced}), it is clear that the exact induced and
equivalent electric surface  currents only deviate from the PO
approximation by a factor of $\frac{1}{\xi}$ for TM mode and
$\frac{1}{\xi^\ast}$ for TE mode.

\begin{figure}
 \hspace{-0.7in}  \includegraphics[width=7.5in, height= 4in]{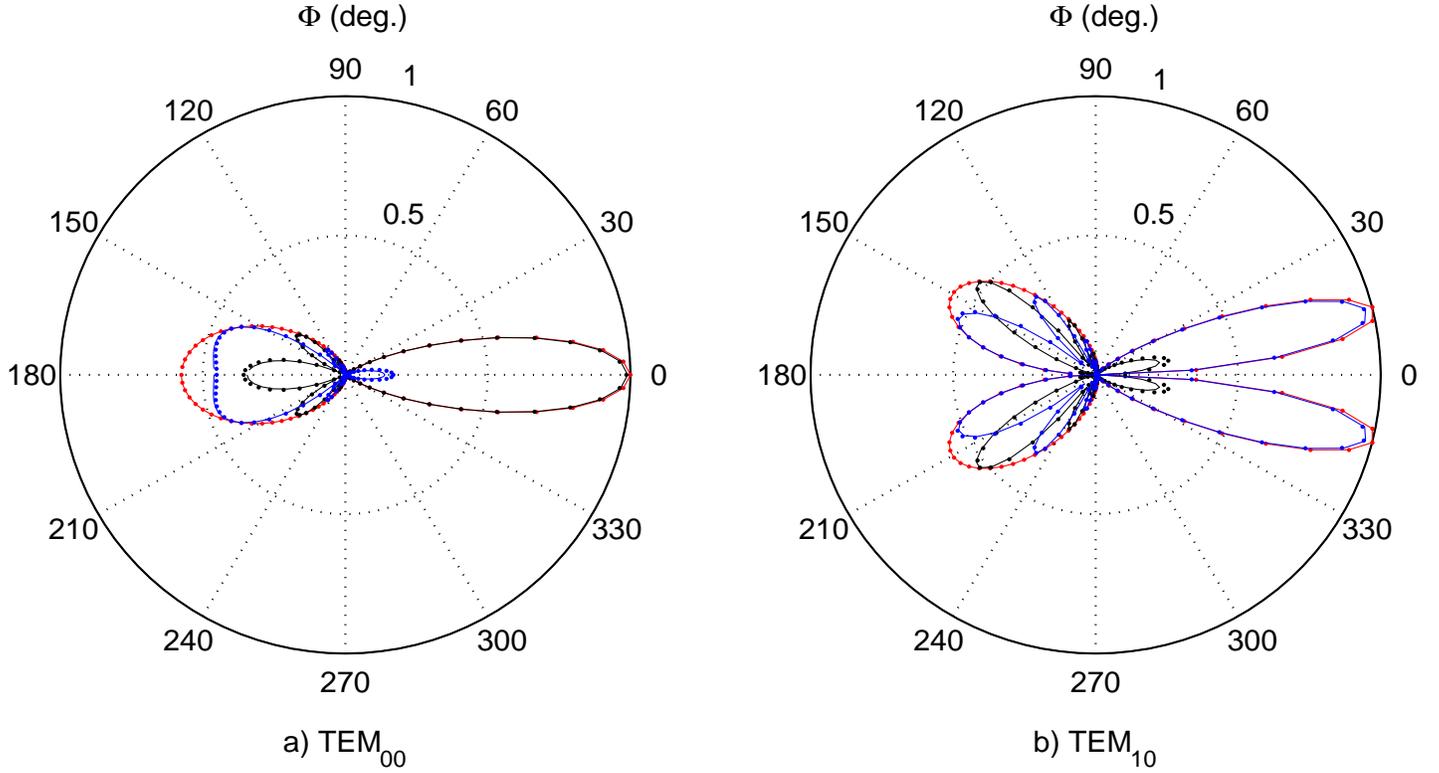}
\caption{ The PO approximation (dots) Vs. result (lines) from MoM:
a) TEM$_{00}$; and b) TEM$_{10}$. Red is for the magnitude; blue is
for the real part; and black is for the imaginary part. Results have
been normalized.
 }
 \label{delTM}
\end{figure}


  \begin{center}
  {\bf IV. Numerical Confirmation:  the Hermite Gaussian Beam }
\end{center}

The incident Hermite Gaussian beam (TEM$_{00}$ and TEM$_{10}$) has
been used to test the result given in (\ref{Js_induced}).  The
TEM$_{mn}$ Hermite Gaussian beam is given as

\vspace*{-0.1in}

{ \small

\begin{eqnarray}\label{Hermite}
  {\bf E}_{mn} = \hat{\bf z} \sqrt{\frac{\eta}{ \pi 2^{m+n-2} m! n! w_y(x)
  w_z(x)
  }}  \hbox{H}_m
  \left( \sqrt{2} \frac{ y }{w_y(x) }\right)
  \hbox{H}_n \left( \sqrt{2} \frac{ z }{w_z(x) }\right)
  \end{eqnarray}

\vspace*{-0.1in}

\begin{eqnarray}\label{Hermite1}
  e^{- \left[y^2\left(  \frac{1}{w_y^2 (x)} + \frac{i k }{2 R_y (x)} \right)
   + z^2\left(  \frac{1}{w_z^2 (x)} + \frac{i k }{2 R_z (x)} \right) \right]}
  e^{  -i  \left[ k x - \left( m+ \frac{1}{2} \right)\arctan\left( \frac{x}{\hbox{\tiny L}_y} \right)
  - \left( n + \frac{1}{2} \right)\arctan\left( \frac{x}{\hbox{\tiny L}_z} \right) \right]},   \nonumber
  \end{eqnarray}

}

\hspace*{-0.22in}where H$_{m,n}$ is the Hermite polynomial and  the
following quantities have been defined,

\vspace*{-0.1in}

{ \small

\begin{eqnarray}\label{def}
 w_\tau(x) = w_{0\tau} \left[ 1 + \frac{x}{\hbox{\small L}_\tau}
 \right]^{\frac{1}{2}}, \ \ \ \ R_\tau(x) = x + \hbox{L}_\tau^2/x, \ \
 \ \ \hbox{L}_\tau = \frac{k w_{0\tau}^2}{2}, \ \ \ \ \tau=y,z
  \end{eqnarray}
}

In our numerical computation, both  TEM$_{00}$ and TEM$_{10}$
Hermite Gaussian beams are   $\hat{\bf z}$-polarized (TM mode only
in the cylindrical coordinate). The symmetrical waist radii have
been set as $w_{0y}= w_{0z}  =1 \lambda$. The radius of the
scattering (diffracting) cylindrical surface is $\rho_0  = 3
\lambda$ and the radius of the observation cylindrical surface is
$\rho = 20 \lambda$.

The   scattered electric field  $ \hbox{E}_z^s$  ($\phi \in
\left[\frac{\pi}{2},\frac{3\pi}{2}\right]$)    and  the  diffracted
electric field $  \hbox{E}_z^d$  ($\phi \in
\left[-\frac{\pi}{2},\frac{\pi}{2}\right]$)   calculated from the PO
approximation have been plotted (dots) in Fig. \ref{delTM}, together
with the result (lines) obtained from the Method of Moment (MoM).
Also, the theoretical induced (equivalent) current given in
(\ref{Js_induced}) has been used to calculate the scattered electric
field $ \hbox{E}_z^s$ and the diffracted electric field $
\hbox{E}_z^d$, which is shown in Fig. \ref{TM} (dots), with good
agreement with the result from the MoM (lines). All plots are for
results on the observation cylindrical surface with radius $\rho =
20 \lambda$.

\begin{figure}
\hspace{-0.7in} \includegraphics[width=7.5in, height=
4in]{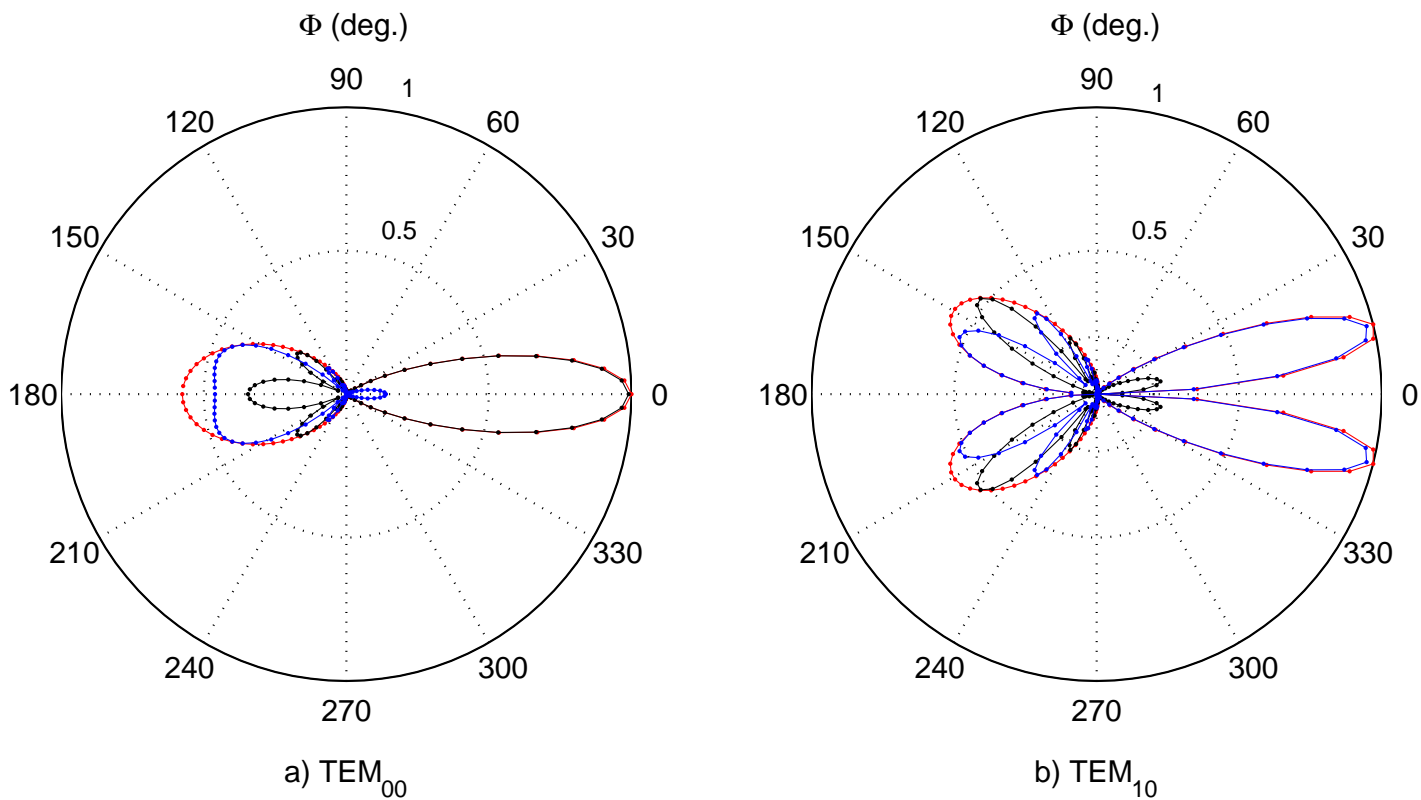}  \caption{ Result (dots) obtained from Eqn.
(\ref{Js_induced}) Vs. result (lines) from MoM: a) TEM$_{00}$ and b)
TEM$_{10}$. Red is for the magnitude; blue is for the real part; and
black is for the imaginary part. Results have been  normalized.
 }
 \label{TM}
 \end{figure}

  \begin{center}
  {\bf Conclusion }
\end{center}

The exact formulas for the induced electric surface current in the
scattering phenomenon and the equivalent electric surface current in
the diffraction phenomenon have been derived, which gives helpful
information of the PO approximation in the cylinder-like surface.

\end{paper}

\end{document}